\documentclass[aps,prd,preprint,preprintnumbers,nofootinbib,longbibliography]{revtex4-2}
\usepackage{amsmath,amssymb,amsfonts}
\usepackage{graphicx}
\usepackage[colorlinks=true, linkcolor=blue, citecolor=magenta, urlcolor=magenta]{hyperref}
\usepackage{bm}
\usepackage{tikz}
\usetikzlibrary{decorations.pathreplacing}
\usepackage{fancyhdr}
\usepackage{etoolbox}

\makeatletter
\patchcmd{\@startsection}
{\@afterindentfalse}
{\@afterindenttrue}
{}{}
\makeatother
\makeatletter
\renewcommand\tableofcontents{%
	\@starttoc{toc}%
}
\makeatother

\makeatletter
\renewcommand\section{\@startsection
	{section}{1}{\z@}%
	{-3.5ex \@plus -1ex \@minus -.2ex}%
	{2.3ex \@plus.2ex}%
	{\centering\normalfont\bfseries}}

\renewcommand\subsection{\@startsection
	{subsection}{2}{\z@}%
	{-3.25ex\@plus -1ex \@minus -.2ex}%
	{1.5ex \@plus .2ex}%
	{\centering\normalfont\bfseries}}
\makeatother
\begin{document}
	\pagestyle{fancy}
	\fancyhf{}
	\fancyhead[R]{\thepage}
	\renewcommand{\headrulewidth}{0pt}
	
	\title{Entanglement entropy between tangent balls in CFT$_D$}
	\author{Jiankun Li}
	\email{lijiankun@stu.scu.edu.cn}
	\affiliation{College of Physics, Sichuan University, Chengdu, 610065, China}
\author{Li Song}
\email{songli1984@scu.edu.cn}
\affiliation{College of Physics, Sichuan University, Chengdu, 610065, China}
	\renewcommand{\abstractname}{}
	
	\begin{abstract}
		We apply the universal method developed in \cite{Jiang:2025jnk} to compute the entanglement entropy  
between two tangent balls in CFT$_D$. 
When taking the radius of one ball to infinity, it gives the entanglement entropy between a ball and its tangent half plane. 
In two-dimensional case, this configuration is equivalent to the entanglement in boundary conformal field theory (BCFT) between the negative half-axis and an interval ending on the boundary.
	\end{abstract}
	
	\maketitle
	\newpage
	\begin{center}
		\textbf{\large CONTENTS}
	\end{center}
	\vspace{-1em}
	
	\tableofcontents
	
	\section{Introduction}
	
	Entanglement entropy has emerged as a fundamental quantity in both quantum gravity and quantum information theory. It is defined by the von Neumann entropy:
	\begin{equation}
		S_{\mathrm{vN}}(A) = -\mathrm{Tr}(\rho_A \log \rho_A),
	\end{equation}
	where \( \rho_A = \mathrm{Tr}_B |\psi_{AB}\rangle \langle \psi_{AB}|\) is the reduced density matrix for subsystem \( A \) in a bipartite system comprising \( A \) and its complement \( B \).
	
\begin{figure}[htbp]
		\centering
		\tikzset{_0jdzjcmgq/.code={
				\pgfsetadditionalshadetransform{
					\pgftransformshift{\pgfpoint{0 bp}{0 bp}}
					\pgftransformscale{1.18}
				}
		}}
		
		\pgfdeclareradialshading{_19m12olk7}{\pgfpoint{0bp}{0bp}}{
			rgb(0bp)=(0.82,0.89,0.97);
			rgb(5.357142857142857bp)=(0.82,0.89,0.97);
			rgb(5.553152901785714bp)=(0.45,0.69,0.91);
			rgb(17.232142857142858bp)=(0.04,0.47,0.84);
			rgb(20.4638671875bp)=(0.33,0.62,0.88);
			rgb(21.696428571428573bp)=(0.53,0.74,0.92);
			rgb(22.8125bp)=(0.04,0.47,0.84);
			rgb(400bp)=(0.04,0.47,0.84)
		}
		
		\tikzset{every picture/.style={line width=0.75pt}} 
		
		\begin{tikzpicture}[x=0.75pt,y=0.75pt,yscale=-1,xscale=1]
			
			\path[shading=_19m12olk7,_0jdzjcmgq]
			(160,149.75) .. controls (160,81.4) and (215.4,26) .. (283.75,26) ..
			controls (352.1,26) and (407.5,81.4) .. (407.5,149.75) ..
			controls (407.5,218.1) and (352.1,273.5) .. (283.75,273.5) ..
			controls (215.4,273.5) and (160,218.1) .. (160,149.75) -- cycle;
			\draw
			(160,149.75) .. controls (160,81.4) and (215.4,26) .. (283.75,26) ..
			controls (352.1,26) and (407.5,81.4) .. (407.5,149.75) ..
			controls (407.5,218.1) and (352.1,273.5) .. (283.75,273.5) ..
			controls (215.4,273.5) and (160,218.1) .. (160,149.75) -- cycle;
			
			\draw[fill=white, fill opacity=1]
			(239.38,149.75) .. controls (239.38,125.24) and (259.24,105.38) .. (283.75,105.38) ..
			controls (308.26,105.38) and (328.13,125.24) .. (328.13,149.75) ..
			controls (328.13,174.26) and (308.26,194.13) .. (283.75,194.13) ..
			controls (259.24,194.13) and (239.38,174.26) .. (239.38,149.75) -- cycle;
			
			\draw[fill={rgb,255:red,126; green,211; blue,33}, fill opacity=1]
			(160,149.75) .. controls (160,142.92) and (177.77,137.38) .. (199.69,137.38) ..
			controls (221.61,137.38) and (239.38,142.92) .. (239.38,149.75) ..
			controls (239.38,156.58) and (221.61,162.13) .. (199.69,162.13) ..
			controls (177.77,162.13) and (160,156.58) .. (160,149.75) -- cycle;
			
			\draw[fill={rgb,255:red,245; green,166; blue,35}, fill opacity=1]
			(328.13,149.75) .. controls (328.13,142.92) and (345.89,137.38) .. (367.81,137.38) ..
			controls (389.73,137.38) and (407.5,142.92) .. (407.5,149.75) ..
			controls (407.5,156.58) and (389.73,162.13) .. (367.81,162.13) ..
			controls (345.89,162.13) and (328.13,156.58) .. (328.13,149.75) -- cycle;
			
			\draw (158.5,79) -- (158.5,46);
			\draw[shift={(158.5,44)}, rotate=90, line width=0.75]
			(10.93,-3.29) .. controls (6.95,-1.4) and (3.31,-0.3) .. (0,0) ..
			controls (3.31,0.3) and (6.95,1.4) .. (10.93,3.29);
			
			\draw (196.69,141.38) node[anchor=north west,inner sep=0.75pt,font=\footnotesize]{\textit{A}};
			\draw (364,143) node[anchor=north west,inner sep=0.75pt,font=\footnotesize]{\textit{B}};
			\draw (161,56.4) node[anchor=north west,inner sep=0.75pt] {$t_{E}$};
			
		\end{tikzpicture}
		\caption{Solid torus (blue region) hosting the CFT. On a time slice, two disjoint entangling regions \(A\) and \(B\) are identified.}
		\label{fig:solid-torus}
	\end{figure}
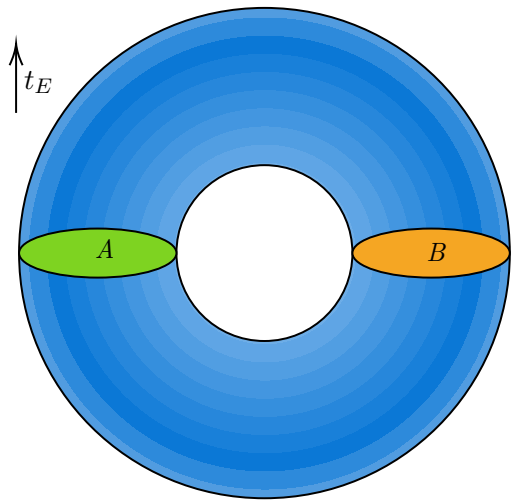

	While the computation of entanglement entropy in CFT$_{2}$ has been well-established via the replica trick \cite{Callan:1994py,Calabrese:2004eu}, extending these results to higher-dimensional theories remains challenging. A universal  approach recently developed in \cite{Jiang:2025jnk} provides a general framework for calculating entanglement entropy of CFTs in arbitrary dimensions by formulating the problem on a solid torus $\mathbb{B}^{D-1} \times S^{1}$, as shown in Figure \ref{fig:solid-torus}. In this formalism, the entanglement entropy $S_\mathrm{disj} (A:B)$ between two disjoint balls 
$A$ and $B$ is expressed in terms of the partition function of the CFT on the solid torus. 
The usual divergent entanglement entropies  $S_\mathrm{adj} (A:B)$ between adjacent entangling regions are
simple  limits of $S_\mathrm{disj} (A:B)$.
The Ryu--Takayanagi (RT) formula \cite{Ryu:2006bv,Ryu:2006ef} emerges naturally from this construction.
	
	Applying this universal approach  to two dimensions naturally reproduces the subtraction method proposed in  \cite{Jiang:2024ijx, Jiang:2025tqu, Jiang:2025dir}, which was recently developed in studies of CFT$_2$. In this special limit, the solid torus geometry reduces to an annulus.

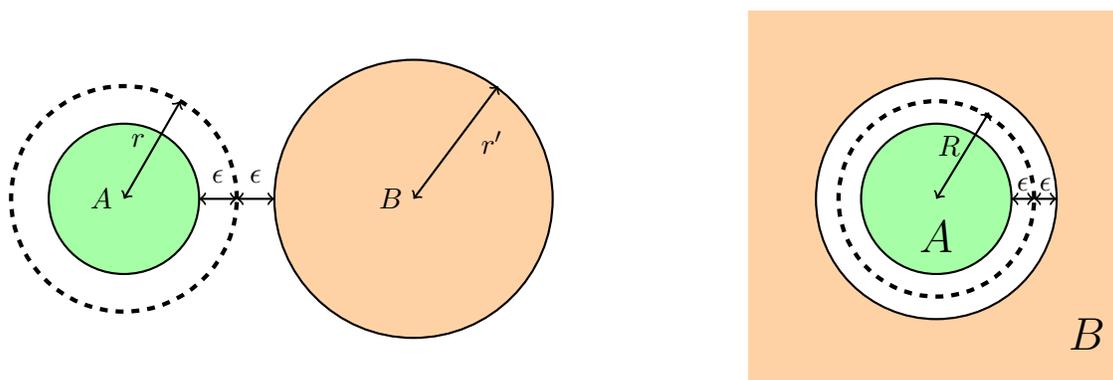
\begin{figure}[htbp]
		\centering
		\begin{tikzpicture}
			\begin{scope}[xshift=-5.4cm]
				\def\R{1}
				\def\r{1.85}
				\def\gap{3.85}
				\draw[dashed, ultra thick] (0,0) circle (1.5);
				\draw[<->, thick] (1,0) -- (1.5,0);
				\node at (1.25,0.3) {$\epsilon$};
				\draw[<->, thick] (1.5,0) -- (2,0);
				\node at (1.75,0.3) {$\epsilon$};
				\def\total{5.5}
				\draw[fill=green!35, thick] (0,0) circle (\R);
				\draw[fill=orange!35, thick] (\gap,0) circle (\r);
				\draw[<->, thick] (0,0) -- (0.75,1.299);
				\node at (0.19,0.77) {$r$};
				\draw[<->, thick] (\gap,0) -- (4.975,1.5);
				\node at (4.89,0.75) {$r'$};
				\node at (-0.3,0) {$A$};
				\node at (3.55,0) {$B$};
			\end{scope}
			\begin{scope}[xshift=5.4cm]
				\def\R{1.0}
				\def\eps{0.6}
				\fill[orange!35] (-2.5,-2.5) rectangle (2.5,2.5);
				\fill[white](0,0)circle(1.6);
				\fill[green!35] (0,0) circle (\R);
				\draw[thick] (0,0) circle (\R);
				\draw[thick] (0,0) circle ({\R+\eps});
				\node at (0,-0.5) {\Large $A$};
				\node at (2,-1.8) {\Large $B$};
				\draw [dashed, ultra thick] (0,0) circle (1.3);
				\draw[<->, thick] (1,0) -- (1.3,0);
				\draw[<->, thick] (1.3,0) -- (1.6,0);
				\node at (1.15,0.2) {$\epsilon$};
				\node at (1.45,0.2) {$\epsilon$};
				\draw[<->, thick] (0,0) -- (0.7,1.15);
				\node at (0.17,0.7) {$R$};
			\end{scope}
		\end{tikzpicture}
		\caption{\small Left panel: Two tangent balls $A$ and $B$ with radii $r$ and $r'$, respectively. The small separation $\epsilon$ between the balls represents a UV cutoff. Right panel: Enclosing configuration of two balls 
$A$ and $B$, separating by a UV cutoff $\epsilon$. Note that the constructions in the left and right panels lead to different divergence structures in the entanglement entropy.}
		\label{fig:$adjacent$}
	\end{figure}

\begin{figure}[htbp]
		\centering
		\begin{tikzpicture}
			\begin{scope}[xshift=-4.4cm]
				\def\R{1}
				\def\gap{3} 
				\def\total{5.5} 
				\draw[fill=green!35, thick] (0,0) circle (\R);
				\node at (0,0) {$A$};
				\fill[orange!35] (2,-2.5) rectangle (5,2.5);
				\node at (3.5,0) {$B$};
				\draw [very thick] (2,-2.5)--(2,2.5);
				
			\end{scope}
			\begin{scope}[xshift=4.4cm]
				\definecolor{lightblue}{rgb}{0.8,0.9,1}
				\fill[lightblue] (-2,-2.5) rectangle (4,2.5);
				\draw[ultra thick, orange!95!black] (-2,0) -- (1,0);
				\draw[ultra thick, green!85!black] (1,0) -- (4,0);
				\fill[white] (1,0) circle (0.2);
				\draw[dashed, ultra thick] (1,0) circle (0.2);
				\node[ultra thick] at (1,-0.5) {\large $\epsilon$};
				\draw[->, thick] (1,0)--(1.178,-0.1);
				\draw[ultra thick, black] (4,-2.5) -- (4,2.5);
				\node[orange!95!black] at (-0.7,-0.5) {$B$};
				\node[green!85!black] at (2.5,-0.5) {$A$};
				\node[black] at (4.4,0) [rotate=270, font=\large]{boundary};
				\node[black] at (-1.6,0.5) {$-\infty$};
				\draw[->, thick] (-1.7,-1.5) -- (-1.2,-1.5) node[below] {$x$};
				\draw[->, thick] (-1.7,-1.5) -- (-1.7,-1.0) node[left] {$\tau$};
			\end{scope}
		\end{tikzpicture}
		\caption{\small The left panel illustrates the entanglement between a ball and a half-space. In CFT$_{2}$, the configuration corresponds to the right panel, which depicts the entanglement between an interval and the negative half-axis.}
		\label{fig:show}
	\end{figure}
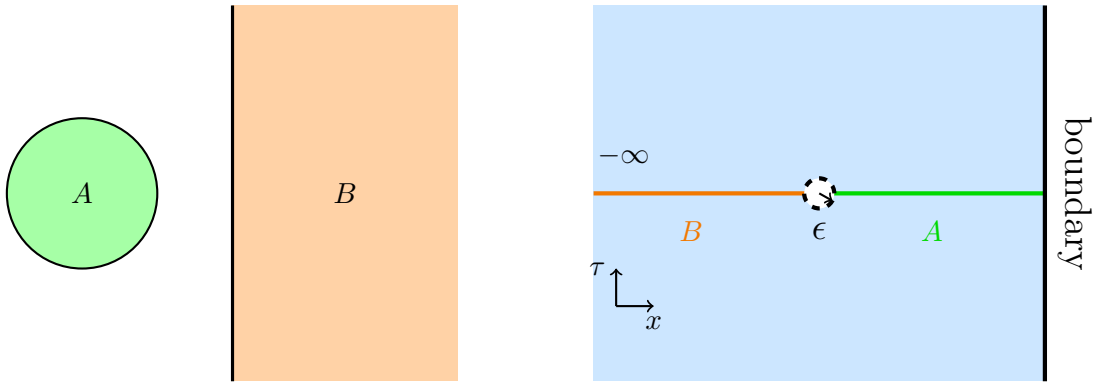

The purpose of this paper is to apply this universal approach in \cite{Jiang:2025jnk} to  specific configurations and provide more verification. 
We first consider the entanglement between two tangent balls, as shown in the left panel of Figure \ref{fig:$adjacent$}. We then compute the entanglement entropy of a ball and a half-space: in the tangent case this is obtained as the special limit where the radius of one ball is sent to infinity, as shown in the left panel of Figure~\ref{fig:show}, while in the nonadjacent configuration the ball is separated from the half-space by a finite distance. In CFT$_{2}$, this configuration corresponds to entanglement in boundary conformal field theory (BCFT) between an interval and the negative half-axis, as depicted in the right panel of Figure \ref{fig:show}.

The structure of the paper is as follows: In Section~\text{II}, we apply the universal method in \cite{Jiang:2025jnk} to the tangent case and further compute the entanglement entropy between a ball and a half-space, which arises as a special case of the tangent configuration. We also compute the entanglement entropy for a configuration in which the ball and the half-space are spatially separated. In Section~\text{III}, we calculate the same entanglement in CFT$_{2}$ by subtraction method \cite{Jiang:2025tqu} and show that two-dimensional solid torus approach coincides with subtraction method.
	
	\section{Entanglement entropy in the tangent configuration and its half-space limit}
	
	In this section, we first apply the universal approach to compute the entanglement entropy in the tangent configuration. We then consider the half-space limit by taking the radius of one ball to infinity and evaluate the corresponding entanglement entropy. Finally, we consider a nonadjacent configuration in which the ball is separated from the half-space by a finite distance and evaluate the corresponding entanglement entropy.
	
	\subsection{Tangent configuration}
	A recent proposal of calculating entanglement entropy for two disjoint regions in all dimensions was developed in  \cite{Jiang:2025jnk}. Consider a CFT in \(D\)-dimensional Euclidean spacetime with the metric $ds_{0}^{2}=dt_{E}^{2}+dy^{2}+\displaystyle \sum_{i=1}^{D-2}dx_{i}^{2}$. The system is defined within a region that is topologically a solid torus, \(\mathcal{B} = \mathbb{B}^{D-1} \times S^1\), where \(\mathbb{B}^{D-1}\) represents a \((D-1)\)-dimensional ball, as depicted in Figure \ref{fig:solid-torus}. Through a Weyl transformation $ds_{0}^{2}\to ds_{0}^{2}/r^{2}$, the metric becomes
	\begin{equation}
		ds^{2}=d\theta^{2}+\left(dr^{2}+\sum_{i=1}^{D-2}dx_{i}^{2} \right)/r^{2},
	\end{equation}
	with the coordinate transformations, $t_{E}=r\sin\theta$, $y=r\cos\theta$. Thus we can regard $\theta$ as imaginary time with period $2\pi$, and evaluate the partition function by treating CFT on $\mathcal{B}$ as a thermal field with the inverse temperature $\beta =2\pi$. By taking a time slice, the entangling region consists of two disjoint \((D-1)\)-balls, \(A\) and \(B\), of the same radius, as shown in Figure \ref{fig:$CFT_{s}$}. The entanglement entropy \(S_\mathrm{disj}(A : B)\) between these two regions can be computed by the replica trick as follows:
	\begin{equation}
		S(n) = \frac{1}{1 - n} \log \left[ \frac{Z_{\mathcal{B}^n}}{Z_\mathcal{B}^n} \right],
	\end{equation}
	where $S(n)$ is the R\'{e}nyi entropy.
\(Z_\mathcal{B}\) and \(Z_{\mathcal{B}^n}\) are the partition functions of the CFT in the original and replicated manifolds, respectively. Here \(\mathcal{B}^n\) is constructed by taking $n$ replicas of \(\mathcal{B}\) and cyclically gluing them along a cut at A (or equivalently B). The metric of \(\mathcal{B}^n\) is
	\begin{equation}
		ds_{(n)}^{2}=n^{2}d\theta^{2}+\left(dr^{2}+\sum_{i=1}^{D-2}dx_{i}^{2} \right)/r^{2}.
	\end{equation}
	By calculating the partition function, the entanglement entropy \(S_\mathrm{disj}(A : B)\) for the disjoint balls is given in  \cite{Jiang:2025jnk} by
	\begin{equation}
		S_\mathrm{disj}(A : B) =\lim_{n\to 1}S(n) = -4\pi \mathcal{E}_{\text{vac}} \text{Vol}(\mathcal{B}^{D-1}),
	\end{equation}
	where \(\mathcal{E}_{\text{vac}}\) is the vacuum (Casimir) energy density.
	
	\begin{figure}[htbp]
		\centering
		\begin{tikzpicture}
			\begin{scope}
				\def\R{1}
				\def\gap{3} 
				\def\total{5.5} 
				\draw[fill=green!35, thick] (0,0) circle (\R);
				\node at (0,0) {$A$};
				\draw[fill=orange!35, thick] (\gap,0) circle (\R);
				\node at (\gap,0) {$B$};
				\draw[<->, thick] (\R,0) -- (\gap-\R,0);
				\node at ({(\gap)/2},0.3) {$2l_1$};
				\draw[<->, thick] (-1,-1.2) -- (4,-1.2);
				\node at ({\gap/2}, -1.6) {$2l_2$};
			\end{scope}
		\end{tikzpicture}
		\caption{\small By considering a time slice of the solid torus in Figure \ref{fig:solid-torus}, one obtains two juxtaposed but disjoint $(D{-}1)$-dimensional balls.}
		\label{fig:$CFT_{s}$}
	\end{figure}
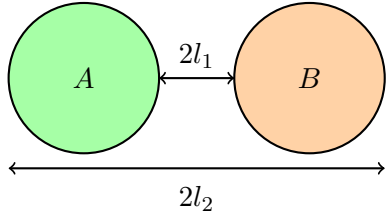
	
	This entanglement entropy is explicitly written as:
	\begin{equation}
		S_\mathrm{disj}(A : B)=- 4\mathcal{E}_{\text{vac}} \pi^{\frac{D}{2}} \frac{\Gamma\left( \frac{D}{2} \right)}{\Gamma(D)} \left( \frac{l_2}{l_1} - 1 \right)^{D-1} {}_2F_1 \left( D - 1, \frac{D}{2}; D; 1 - \frac{l_2}{l_1} \right),
	\end{equation}
	where \(l_1\) and \(l_2\) are marked in Figure \ref{fig:$CFT_{s}$}. 
For clarity, we present the explicit expressions in various dimensions: 
	\begin{align}
		D = 2 &: \quad 4\pi \mathcal{E}_2 \log \frac{l_2}{l_1}, \nonumber \\
		D = 3 &: \quad 4\pi^{2} \mathcal{E}_3 \left( \sqrt{\frac{l_2}{l_1}} + \sqrt{\frac{l_1}{l_2}} \right) - 8 \pi^{2} \mathcal{E}_3, \nonumber \\
		D = 4 &: \quad 2\pi^{2} \mathcal{E}_4 \left( \frac{l_2}{l_1} - \frac{l_1}{l_2} \right) - 4 \pi^{2} \mathcal{E}_4 \log \frac{l_2}{l_1}, \nonumber \\
		D = 5 &: \quad \frac{\pi^{3} \mathcal{E}_5}{3} \left[ \left( \frac{l_2}{l_1} \right)^{3/2} + \left( \frac{l_1}{l_2} \right)^{3/2}
		- 9 \left( \sqrt{\frac{l_2}{l_1}} + \sqrt{\frac{l_1}{l_2}} \right) \right] + \frac{16 \pi^{3} \mathcal{E}_5}{3}, \nonumber \\
		D = 6 &: \quad \frac{\pi^{3} \mathcal{E}_6}{6} \left[ \left( \frac{l_2}{l_1} \right)^{2} - \left( \frac{l_1}{l_2} \right)^{2} - 8 \left( \frac{l_2}{l_1} - \frac{l_1}{l_2} \right) \right]
		+ 2 \pi^{3} \mathcal{E}_6 \log \frac{l_2}{l_1},
		\label{several expressions}
	\end{align}
where	we adopted the convention $\mathcal{E}_{D} = - \mathcal{E}_{\mathrm{vac}}$ which takes different values in different dimensions.
	
Note the entanglement entropy above is computed under the conditions that $A$ and $B$ are symmetric.	To get the results for general configurations, it is convenient to use a conformal invariant to describe the geometric configuration of the system. Specifically, we make use of the inversive product \(\varrho\) \cite{beardon2012geometry} of two spheres \(\Sigma(\vec{x}, r)\) and \(\Sigma(\vec{x}', r')\), which is defined as:
	\begin{equation}
		\varrho = \left| \frac{r^2 + r'^2 - |\vec{x} - \vec{x}'|^2}{2rr'} \right|.
	\end{equation}
	Here, \(\Sigma(\vec{x}, r)\) denotes a sphere located at \(\vec{x}\) with radius \(r\). Accordingly, $\varrho$ is invariant under global conformal transformation. Configurations sharing the same $\varrho$ are conformally equivalent and have identical entanglement properties. Thus, the entanglement entropy can be written in terms of inversive product $\varrho$:
	\begin{equation}
		S_\mathrm{disj}(A:B) = -4\mathcal{E}_{\text{vac}} \pi^{\frac{D}{2}} \frac{\Gamma\left( \frac{D}{2} \right)}{\Gamma(D)} \left( \frac{2\sqrt{2}}{\sqrt{\varrho + 1} - \sqrt{2}} \right)^{D-1} {}_2F_1\left( D-1, \frac{D}{2}; D; \frac{2\sqrt{2}}{\sqrt{2} - \sqrt{\varrho + 1}} \right).
		\label{inverse expression}
	\end{equation}
	
	In this paper, we analyze the entanglement of two balls in the tangent limit, as illustrated in the left panel of Figure \ref{fig:$adjacent$}. For clarity, we denote the radii of $A$ and $B$ by $r$ and $r'$ respectively. The two balls are separated by a UV cutoff $\epsilon$. The inversive product is
	\begin{equation}
		\varrho=\frac{2rr'+4r\epsilon+2r'\epsilon}{2r'\left(r-\epsilon\right)}.
	\end{equation}
	Thus, from Equation (\ref{inverse expression}), the entanglement entropy is
	\begin{equation}
		S_\mathrm{disj}(A:B)  = -4\mathcal{E}_{\text{vac}} \pi^{\frac{D}{2}} \frac{\Gamma\left( \frac{D}{2} \right)}{\Gamma(D)} \left( \frac{2}{\sqrt{\frac{r\left(r'+\epsilon\right)}{r'\left(r-\epsilon\right)}} - 1} \right)^{D-1} {}_2F_1\left( D-1, \frac{D}{2}; D; \frac{2}{1-\sqrt{\frac{r\left(r'+\epsilon\right)}{r'\left(r-\epsilon\right)}}} \right).
	\end{equation}
	Concrete expressions for this equation in specific dimensions are shown as follows:
	\begin{align}
		D=2 &: \quad 4\pi \mathcal{E}_2 \log \frac{2rr'+\left(r-r'\right)\epsilon+2\sqrt{rr'\left(r'+\epsilon\right)\left(r-\epsilon\right)}}{\left(r+r'\right)\epsilon},   \nonumber \\
		D=3 &: \quad 8\pi^{2} \mathcal{E}_3 \left(\sqrt{\frac{r\left(r'+\epsilon\right)}{\left(r+r'\right)\epsilon}}-1\right), \nonumber \\
		D=4 &: \quad 4\pi^{2} \mathcal{E}_4 \left(\frac{2\sqrt{rr'\left(r'+\epsilon\right)\left(r-\epsilon\right)}}{\left(r+r'\right)\epsilon}-\log \frac{2rr'+\left(r-r'\right)\epsilon+2\sqrt{rr'\left(r'+\epsilon\right)\left(r-\epsilon\right)}}{\left(r+r'\right)\epsilon}\right), \nonumber \\
		D=5 &: \quad \frac{\pi^{3} \mathcal{E}_5}{3}\left[8\left(\frac{r\left(r'+\epsilon\right)}{\left(r+r'\right)\epsilon}\right)^{\frac{3}{2}}-24\sqrt{\frac{r\left(r'+\epsilon\right)}{\left(r+r'\right)\epsilon}}+16\right],  \nonumber \\
		D=6 &: \quad \frac{\pi^{3}\mathcal{E}_6}{3}
		\left(
		\frac{4\sqrt{\,rr'\left(r'+\epsilon\right)\left(r-\epsilon\right)} \left(2rr'-3r\epsilon-5r'\epsilon\right)}{\left(r+r'\right)^{2}\epsilon^{2}} \right. \nonumber
		\\[6pt]
		&\quad \quad\quad\quad\quad\quad\quad\quad \left. + 6\log
		\frac{2rr'+\left(r-r'\right)\epsilon+2\sqrt{rr'\left(r'+\epsilon\right)\left(r-\epsilon\right)}}{\left(r+r'\right)\epsilon} \right). 
		\label{tangent}
	\end{align}
After taking $\epsilon\to 0$, one get the usual divergent entanglement entropy between two tangent balls:
\begin{align}
	D=2 &: \quad 4\pi \mathcal{E}_2 \log 
	\frac{4rr'}{\left(r+r'\right)\epsilon}+\mathcal{O}\left(\epsilon\right),   \nonumber \\
	D=3 &: \quad 8\pi^{2} \mathcal{E}_3 \left(\sqrt{\frac{rr'}{\left(r+r'\right)\epsilon}}-1\right)+\mathcal{O}\left(\sqrt{\epsilon}\right), \nonumber \\
	D=4 &: \quad 4\pi^{2} \mathcal{E}_4 \left(\frac{2rr'}{\left(r+r'\right)\epsilon}+\frac{r-r'}{r+r'}-\log\frac{4rr'}{\left(r+r'\right)\epsilon}\right)+\mathcal{O}\left(\epsilon\right), \nonumber \\
	D=5 &: \quad \frac{\pi^{3} \mathcal{E}_5}{3}\left[8\left(\frac{rr'}{\left(r+r'\right)\epsilon}\right)^{\frac{3}{2}}-24\sqrt{\frac{rr'}{\left(r+r'\right)\epsilon}}+16\right]+\mathcal{O}\left(\sqrt{\epsilon}\right),  \nonumber \\
	D=6 &: \quad \frac{\pi^{3} \mathcal{E}_6}{3} \left(8\left(\frac{rr'}{\left(r+r'\right)\epsilon}\right)^{2}-\frac{8rr'\left(r+3r'\right)}{\left(r+r'\right)^{2}\epsilon}+\frac{-7r^{2}-6rr'+9r'^{2}}{\left(r+r'\right)^{2}}+6\log \frac{4rr'}{\left(r+r'\right)\epsilon}\right)+\mathcal{O}\left(\epsilon\right).
\end{align}

	\subsection{Half-space limit}
\begin{figure}[htbp]
		\begin{tikzpicture}
			\centering
			\begin{scope}[xshift=-4cm]
				\def\R{1}
				\def\gap{3} 
				\def\total{5.5} 
				\draw[fill=green!35, thick] (0,0) circle (\R);
				\node at (-0.3,0) {$A$};
				\fill[orange!35] (2,-2.5) rectangle (5,2.5);
				\node at (3.5,0) {$B$};
				\draw [very thick] (2,-2.5)--(2,2.5);
				\draw[<->, thick] (0,0) -- (0.5,0.89);
				\node at (0.47,0.37) {$R$};
				\draw[<->, thick] (1,0) -- (2,0);
				\node at (1.5,0.3) {$d$};
				
			\end{scope}
			\begin{scope}[xshift=4cm]
				\def\R{1}
				\def\gap{3} 
				\def\total{5.5} 
				\draw[fill=green!35, thick] (0,0) circle (\R);
				\draw[dashed, ultra thick] (0,0) circle (1.5);
				\node at (-0.3,0) {$A$};
				\fill[orange!35] (2,-2.5) rectangle (5,2.5);
				\node at (3.5,0) {$B$};
				\draw [very thick] (2,-2.5)--(2,2.5);
				\draw[<->, thick] (1,0) -- (1.5,0);
				\node at (1.25,0.3) {$\epsilon$};
				\draw[<->, thick] (1.5,0) -- (2,0);
				\node at (1.75,0.3) {$\epsilon$};
				\draw[<->, thick] (0,0) -- (0.75,1.299);
				\node at (0.19,0.77) {$R$};
			\end{scope}
		\end{tikzpicture}
		\caption{\small The left panel illustrates the entanglement between a ball and a half-space. The spherical region \(A\) is disjoint with region \(B\). 
			By taking the adjacent limit, as shown by the right panel, the inversive product is defined by slightly shifting the radius by \(\pm \epsilon\).}
		\label{fig:halfspace-ball}
	\end{figure}
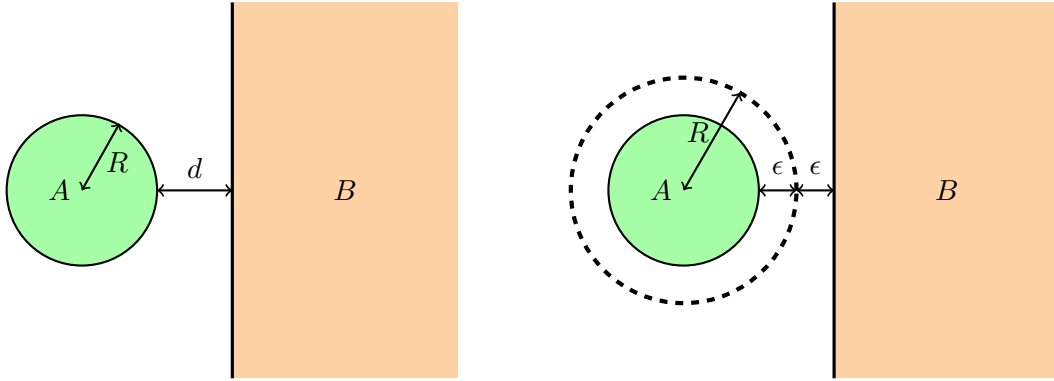

	As a special case of the tangent configuration, we analyze the entanglement behavior in the limit $r'\to\infty$. Consider the half-space \(\mathcal{H}=\{\vec{x}: \vec{n} \cdot (\vec{x}-\vec{x}_{0}) \geq 0 \}\), where \(\vec{n}\) is the normal vector and \(\vec{x}_{0}\) is a point on its boundary. We define a ball contained in the half-space \(\mathcal{H}\) with radius \(r'\) and tangent to its boundary at point \(\vec{x}_{0}\) as \(\mathcal{B}(\vec{x}, r')=\{\vec{x}: \left\| \vec{x}-(\vec{x}_{0}+r'\vec{n}) \right\|^{2} \leq r'^{2}\}\). In the limit where the radius of a \((D-1)\) dimensional ball tends to infinity, we obtain \(\lim\limits_{r'\to\infty}\mathcal{B}(\vec{x}, r')=\{\vec{x}:\vec{n} \cdot (\vec{x}-\vec{x}_{0}) \geq 0 \}\), which is exactly \(\mathcal{H}\). Thus, the entanglement between a finite ball and a ball of infinite radius is equivalent to that between a ball and the half-space. Considering the entangling surfaces shown in Figure \ref{fig:halfspace-ball}, the inversive product \(\varrho\) \cite{beardon2012geometry} for the boundary sphere \(\Sigma(\vec{x}, r)\) and the boundary plane \(P(\vec{n},0)\) is
	\begin{equation}
		\varrho=\frac{R+\epsilon}{R-\epsilon}.
	\end{equation}
	The entanglement entropy is then:
	\begin{equation}
		S_\mathrm{disj}(A:B)=-4\pi^{\frac{D}{2}}\mathcal{E}_{vac}\frac{\Gamma(\frac{D}{2})}{\Gamma(D)}(\frac{4R}{\epsilon}-3)^{D-1}{}_2F_1(D-1,\frac{D}{2},D,3-\frac{4R}{\epsilon}).
	\end{equation}
For various dimensions, taking the limit $\epsilon\to 0$, the usual divergent entanglement entropy between a pair of  tangent 
ball  and  half-space is 

	\begin{align}
		D = 2 &: \quad 4\pi \mathcal{E}_2 \log \frac{4R}{\epsilon} +\mathcal{O}(\frac{\epsilon}{R}), \nonumber \\
		D = 3 &: \quad 8\pi^{2} \mathcal{E}_3 \left( \sqrt{\frac{R}{\epsilon}} -1 \right), \nonumber \\
		D = 4 &: \quad 4\pi^{2} \mathcal{E}_4 \left( \frac{2R}{\epsilon} -1 - \log \frac{4R}{\epsilon} \right) +\mathcal{O}(\frac{\epsilon}{R}) , \nonumber \\
		D = 5 &: \quad \frac{\pi^{3} \mathcal{E}_5}{3} \left[ 8 \left( \frac{R}{\epsilon} \right)^{3/2} - 24 \sqrt{\frac{R}{\epsilon}} +16  \right], \nonumber \\
		D = 6 &: \quad \frac{\pi^{3} \mathcal{E}_6}{3} \left[ 8 \left( \frac{R}{\epsilon} \right)^{2} - \frac{24R}{\epsilon} +9 + 6 \log \frac{4R}{\epsilon} \right]
		+\mathcal{O}(\frac{\epsilon}{R}).
		\label{adjacent}
	\end{align}
Equation (\ref{adjacent}) precisely corresponds to the limiting case of equation (\ref{tangent}) under $r'\to\infty$ and $\epsilon\to 0$. Notably, in the adjacent limit, the expression of the tangent configuration differs from that of the enclosing configuration, as shown in the right panel of Figure \ref{fig:$adjacent$}. Consequently, the divergence structure of the entanglement entropy in this setup is different from the two-sphere case discussed in \cite{Jiang:2025jnk}.
	
	For another simpler case, the entanglement of two half-spaces with two cutoffs $l_{1}=\epsilon\to0$ and $l_{2}=2\xi\to\infty$, we can easily derive that the entanglement entropy is:
	\begin{align}
		D = 2 &: \quad 4\pi \mathcal{E}_2 \log \frac{2\xi}{\epsilon}, \nonumber \\
		D = 3 &: \quad 4\pi^{2} \mathcal{E}_3 \left( \sqrt{\frac{2\xi}{\epsilon}} -2 \right) + \mathcal{O}(\sqrt{\frac{\epsilon}{\xi}}), \nonumber \\
		D = 4 &: \quad 4\pi^{2} \mathcal{E}_4 \left( \frac{\xi}{\epsilon} - \log \frac{2\xi}{\epsilon} \right) + \mathcal{O}(\frac{\epsilon}{\xi}), \nonumber \\
		D = 5 &: \quad \frac{\pi^{3} \mathcal{E}_5}{3} \left[ \left( \frac{2\xi}{\epsilon} \right)^{3/2} - 9  \sqrt{\frac{2\xi}{\epsilon}} +16 \right] + \mathcal{O}(\sqrt{\frac{\epsilon}{\xi}}) , \nonumber \\
		D = 6 &: \quad \frac{2\pi^{3} \mathcal{E}_6}{3} \left[ \left( \frac{\xi}{\epsilon} \right)^{2} -\frac{4\xi}{\epsilon} + 3\log \frac{2\xi}{\epsilon} \right] + \mathcal{O}(\frac{\epsilon}{\xi}),
	\end{align}
which represent the entanglement between the left and right halves.

    \subsection{Nonadjacent half-space limit}
    For completeness, we also discuss the half-space limit in the nonadjacent configuration, shown in the left panel of Figure \ref{fig:halfspace-ball}. In this setup, the ball and the half-space are separated by a finite distance $d$. The inversive product is
    \begin{equation}
    	\varrho=\frac{\left|\vec{n}\cdot\vec{x}\right|}{r}=\frac{R+d}{R}.
    \end{equation}
    Thus the entanglement entropy takes the form
    \begin{equation}
    	S_\mathrm{disj}(A : B)=- 4\mathcal{E}_{\text{vac}} \pi^{\frac{D}{2}} \frac{\Gamma\left( \frac{D}{2} \right)}{\Gamma(D)} \left( \frac{4R}{d} + 2\sqrt{\frac{4R^{2}}{d^{2}}+\frac{2R}{d}} \right)^{D-1} {}_2F_1 \left( D - 1, \frac{D}{2}; D; -\frac{4R}{d} - 2\sqrt{\frac{4R^{2}}{d^{2}}+\frac{2R}{d}} \right).
    \end{equation}
    We also record the explicit expressions in several dimensions:
    \begin{align}
    	D = 2 &: \quad 8\pi \mathcal{E}_2 \log \left(\sqrt{\frac{2R}{d}} + \sqrt{1+\frac{2R}{d}}\right), \nonumber \\
    	D = 3 &: \quad 4\pi^{2} \mathcal{E}_3 \frac{\left(\sqrt{\frac{2R}{d}}+\sqrt{1+\frac{2R}{d}}-1\right)^{2}}{\sqrt{\frac{2R}{d}}+\sqrt{1+\frac{2R}{d}}}, \nonumber \\
    	D = 4 &: \quad 2\pi^{2}\mathcal{E}_{4}\left[\left(\sqrt{\frac{2R}{d}} + \sqrt{1+\frac{2R}{d}}\right)^{2}- \frac{1}{\left(\sqrt{\frac{2R}{d}} + \sqrt{1+\frac{2R}{d}}\right)^{2}}-4\log\left(\sqrt{\frac{2R}{d}} + \sqrt{1+\frac{2R}{d}}\right)\right]
    	, \nonumber \\
    	D = 5 &: \quad \frac{\pi^{3}\mathcal{E}_{5}}{3}
    	\left[\left(\sqrt{\frac{2R}{d}} + \sqrt{1+\frac{2R}{d}}\right)^{3}+ \frac{1}{\left(\sqrt{\frac{2R}{d}} + \sqrt{1+\frac{2R}{d}}\right)^{3}}
    	\right.\nonumber \\[6pt]
    	&\quad\quad\quad\quad\quad\quad\quad\quad\quad\quad\quad\quad\quad \left.- 9 \left(\sqrt{\frac{2R}{d}} + \sqrt{1+\frac{2R}{d}}+ \frac{1}{\sqrt{\frac{2R}{d}} + \sqrt{1+\frac{2R}{d}}}\right)+ 16\right], \nonumber \\
    	D = 6 &: \quad \frac{\pi^{3}\mathcal{E}_{6}}{6}\left[\left(\sqrt{\frac{2R}{d}} + \sqrt{1+\frac{2R}{d}}\right)^{4}- \frac{1}{\left(\sqrt{\frac{2R}{d}} + \sqrt{1+\frac{2R}{d}}\right)^{4}} \right. \nonumber\\
    	&\left. -8 \left(\left(\sqrt{\frac{2R}{d}} + \sqrt{1+\frac{2R}{d}}\right)^{2}- \frac{1}{\left(\sqrt{\frac{2R}{d}} + \sqrt{1+\frac{2R}{d}}\right)^{2}}\right)+24\log\left(\sqrt{\frac{2R}{d}} + \sqrt{1+\frac{2R}{d}}\right)\right].
    \end{align}
    From this expression, one can readily see that the entanglement entropy in the nonadjacent configuration is finite. In particular, for D=2, the expression is
    \begin{equation}
    	S = 8\pi \mathcal{E}_2 \log \left(\sqrt{\frac{2R}{d}} + \sqrt{1+\frac{2R}{d}}\right)= 4\pi \mathcal{E}_2 \log \left(1+\frac{4R}{d}+2\sqrt{\frac{2R}{d}\left(\frac{2R}{d}+1\right)}\right).
    	\label{eq: non-tangent}
    \end{equation}
	This expression precisely matches the entanglement entropy of a quantum state with an interval of length $2R$ and a semi-infinite line that is separated from it by a distance $d$, as will be illustrated in the next section. By treating the finite separation $d$ as an infinitesimal UV regulator and sending $d\to0$, the nonadjacent result smoothly reduces to the tangent one.
	
	\section{A specific example of computation in CFT$_2$}
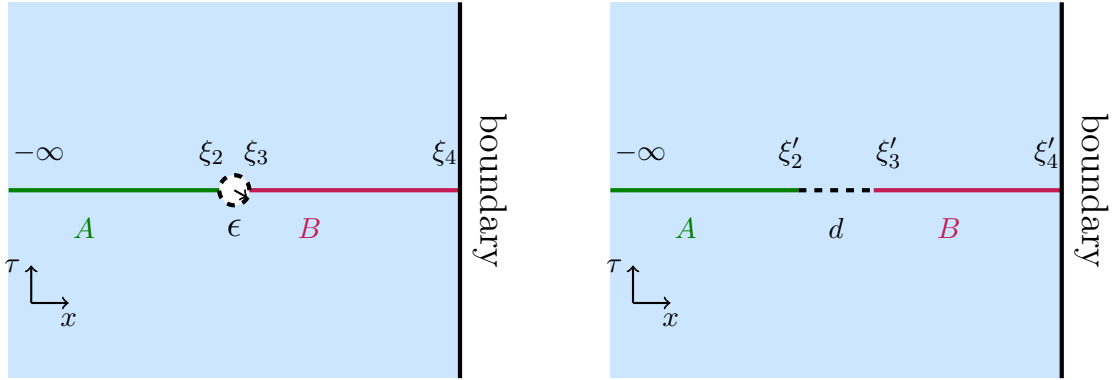
\begin{figure}[htbp]
		\begin{tikzpicture}
			\centering
			\begin{scope}[xshift=-4cm]
				\definecolor{lightblue}{rgb}{0.8,0.9,1}
				\fill[lightblue] (-2,-2.5) rectangle (4,2.5);
				\draw[ultra thick, green!50!black] (-2,0) -- (1,0);
				\draw[ultra thick, purple!90] (1,0) -- (4,0);
				\fill[white] (1,0) circle (0.2);
				\draw[dashed, ultra thick] (1,0) circle (0.2);
				\node[ultra thick] at (1,-0.5) {\large $\epsilon$};
				\draw[->, thick] (1,0)--(1.178,-0.1);
				\draw[ultra thick, black] (4,-2.5) -- (4,2.5);
				\node[green!50!black] at (-1,-0.5) {$A$};
				\node[purple!90] at (2,-0.5) {$B$};
				\node[black] at (4.4,0) [rotate=270, font=\large]{boundary};
				\node[black] at (-1.6,0.5) {$-\infty$};
				\node[black] at (0.7,0.5) {$\xi_{2}$};
				\node[black] at (1.3,0.5) {$\xi_{3}$};
				\node[black] at (3.8,0.5) {$\xi_{4}$};
				\draw[->, thick] (-1.7,-1.5) -- (-1.2,-1.5) node[below] {$x$};
				\draw[->, thick] (-1.7,-1.5) -- (-1.7,-1.0) node[left] {$\tau$};
			\end{scope}
			\begin{scope}[xshift=4cm]
				\definecolor{lightblue}{rgb}{0.8,0.9,1}
				\fill[lightblue] (-2,-2.5) rectangle (4,2.5);
				\draw[ultra thick, green!50!black] (-2,0) -- (0.5,0);
				\draw[ultra thick, purple!90] (1.5,0) -- (4,0);
				\draw[ultra thick, black] (4,-2.5) -- (4,2.5);
				\node[green!50!black] at (-1,-0.5) {$A$};
				\node[purple!90] at (2.5,-0.5) {$B$};
				\draw[dashed, ultra thick] (0.5,0) -- (1.5,0);
				\node[black] at (1,-0.5) {$d$};
				\node[black] at (4.4,0) [rotate=270, font=\large]{boundary};
				\node[black] at (-1.6,0.5) {$-\infty$};
				\node[black] at (0.4,0.5) {$\xi_{2}'$};
				\node[black] at (1.7,0.5) {$\xi_{3}'$};
				\node[black] at (3.8,0.5) {$\xi_{4}'$};
				\draw[->, thick] (-1.7,-1.5) -- (-1.2,-1.5) node[below] {$x$};
				\draw[->, thick] (-1.7,-1.5) -- (-1.7,-1.0) node[left] {$\tau$};
			\end{scope}
		\end{tikzpicture}
		\caption{\small This picture shows the half-space limit case in $\mathrm{CFT}_{2}$. In the left panel, region A is the negative half-axis ($-\infty, \xi_{2}$), while region B is the finite interval terminating on the boundary ($\xi_{3}, \xi_{4}$). The two intervals are separated by a small distance $2\epsilon=\xi_{3}-\xi_{2}$, which serves as a UV regulator. In the right panel, region A and region B are separated by a finite distance $d$.}
		\label{fig:$BCFT$}
	\end{figure}

	We have considered all dimensional cases in solid torus CFT and successfully obtained BCFT results. For $\mathrm{CFT_{2}}$, the vacuum energy is related to the central charge \cite{DiFrancesco:1997nk}:
	\begin{equation}
		\mathcal{E}_{vac}=-\frac{c}{24\pi},
	\end{equation}
	and the usual divergent adjacent entanglement entropy is
	\begin{equation}
		S_\mathrm{adj}=\frac{c}{6}\log\frac{4R}{\epsilon},
\label{eq: half}
	\end{equation}
	which precisely corresponds to the previous results of entanglement entropy for an interval with length $2R$ in the half-space. Moreover, the nonadjacent entanglement entropy in equation (\ref{eq: non-tangent}) turns to
	\begin{equation}
		S =  \frac{c}{6} \log \left(1+\frac{4R}{d}+2\sqrt{\frac{2R}{d}\left(\frac{2R}{d}+1\right)}\right).
\label{eq: nonadjacent half}
	\end{equation}
	This result also precisely matches the entanglement entropy for an interval with length $2R$ and the negative half-axis separated by a distance $d$, as shown in the right panel of Figure \ref{fig:$BCFT$}.
	
	In fact, these results can be obtained by subtraction method in a similar way. As \cite{Jiang:2024ijx, Jiang:2025tqu, Jiang:2025dir} have illustrated, for two disjoint intervals in CFT$_{2}$ at the time slice \( \tau = 0 \), we can remove two discs from the Euclidean path integral region to construct a pure state \( \psi_{AB} \). The doubly connected region can then be conformally mapped onto an annulus with width \( W = \log \frac{r_2}{r_1} \) and two boundary states \( |a,b\rangle \), as illustrated in the right panel of Figure \ref{fig:$CFT_{2}$}. The entanglement entropy in the annulus is defined in terms of the R\'{e}nyi entropy:
	\begin{equation}
		S^{(n)}(A) = \frac{c}{12} \left( 1 + \frac{1}{n} \right) W + g_a + g_b.
	\end{equation}
	
	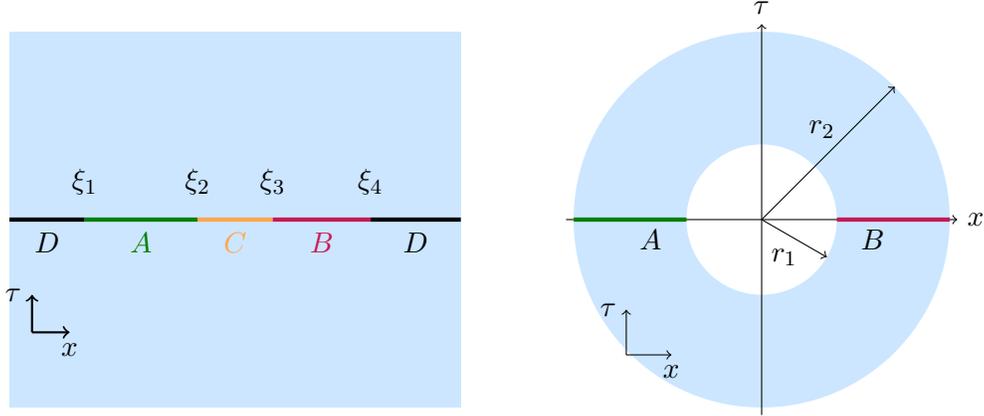
\begin{figure}[htbp]
		\centering
		\begin{tikzpicture}
			\begin{scope}[xshift=-4cm]
				\definecolor{lightblue}{rgb}{0.8,0.9,1}
				\fill[lightblue] (-2,-2.5) rectangle (4,2.5);
				\draw[ultra thick, black] (-2,0) -- (-1,0);
				\draw[ultra thick, green!50!black] (-1,0) -- (0.5,0);
				\draw[ultra thick, orange!70] (0.5,0) -- (1.5,0);
				\draw[ultra thick, purple!90] (1.5,0) -- (2.8,0);
				\draw[ultra thick, black] (2.8,0) -- (4,0);
				\node[black] at (-1.5,-0.3) {$D$};
				\node[green!50!black] at (-0.25,-0.3) {$A$};
				\node[orange!70] at (1,-0.3) {$C$};
				\node[purple!90] at (2.15,-0.3) {$B$};
				\node[black] at (3.4,-0.3) {$D$};
				\draw[->, thick] (-1.7,-1.5) -- (-1.2,-1.5) node[below] {$x$};
				\draw[->, thick] (-1.7,-1.5) -- (-1.7,-1.0) node[left] {$\tau$};
				\node[black] at (-1,0.5) {$\xi_{1}$};
				\node[black] at (0.5,0.5) {$\xi_{2}$};
				\node[black] at (1.5,0.5) {$\xi_{3}$};
				\node[black] at (2.8,0.5) {$\xi_{4}$};
			\end{scope}
			\begin{scope}[xshift=4cm]
				\definecolor{lightblue}{rgb}{0.8,0.9,1}
				\fill[lightblue] (0,0) circle (2.5);
				\fill[white] (0,0) circle (1); 
				\node[black] at (190:1.5) {$A$};
				\node[black] at (-10:1.5) {$B$};
				\draw[->] (-2.6,0) -- (2.6,0) node[right] {$x$};
				\draw[->] (0,-2.6) -- (0,2.6) node[above] {$\tau$};
				\draw[->] (-1.8,-1.8) -- (-1.2,-1.8) node[below] {$x$};
				\draw[->] (-1.8,-1.8) -- (-1.8,-1.2) node[left] {$\tau$};
				\draw[green!50!black, ultra thick] (180:1) -- (180:2.5);
				\draw[purple!90, ultra thick] (0:1) -- (0:2.5);
				\draw[->] (0,0) -- (-30:1);
				\draw[->] (0,0) -- (45:2.5);
				\node at (0.8,1.2){$r_{2}$};
				\node at (0.3,-0.5){$r_{1}$};
			\end{scope}
		\end{tikzpicture}
		\caption{\small The left panel illustrates two disjoint intervals in $\mathrm{CFT}_2$. By subtracting segments $C$ and $D$, one obtains an annular region that represents a pure state $\psi_{AB}$, as shown in the right panel.}
		\label{fig:$CFT_{2}$}
	\end{figure}
	
	Here, \( g_{a,b} = \log \langle a, b | 0 \rangle \) represent the Affleck-Ludwig boundary entropies \cite{Affleck:1991tk}, which encode the information about the undetectable regions and are  irrelevant under the large $c$ limit. 
Consequently, the universal term of entanglement entropy between subsystems \(A\) and \(B\) is given as:
	\begin{equation}
		S_{\mathrm{vN}}(A : B) = \lim_{n \to 1} S^{(n)}(A) = \frac{c}{6} W.
	\end{equation}
	For two intervals with endpoints \(\xi_1, \xi_2, \xi_3, \xi_4\), a conformally invariant cross ratio is defined as:
	\begin{equation}
		\eta = \frac{(\xi_2 - \xi_1)(\xi_4 - \xi_3)}{(\xi_3 - \xi_2)(\xi_4 - \xi_1)},
	\end{equation}
	and the entanglement entropy can be written as:
	\begin{align}
		S_{\text{vN}}(A : B) &= \frac{c}{6} W \notag \\
		&= \frac{c}{6} \log \left[ 1 + 2\eta + 2\sqrt{\eta(\eta+1)} \right].
		\label{2D EE}
	\end{align}
	
	To reduce the general case to the half-space BCFT, we consider the negative half-axis (region A) and an interval (region B) by setting \( \xi_1 = -\infty \), \( \xi_2=-2R-\epsilon \), \( \xi_3=-2R+\epsilon \), and \( \xi_4 =- \epsilon \to 0 \), as shown in the left panel of Figure \ref{fig:$BCFT$}. The cross ratio is given by \( \eta = \frac{R}{\epsilon} \). Substituting equation (\ref{2D EE}) into the expression for \( S_{\text{vN}}(A : B) \), we obtain the following result:
	
	\begin{equation}
		S_{\text{vN}}(A : B) = \frac{c}{6} \log \frac{4R}{\epsilon},
	\end{equation}
	which is in agreement with equation (\ref{eq: half}), the half-space entanglement entropy with boundaries. In the AdS/CFT correspondence, the entanglement entropy of the half-space corresponds to half of the geodesic cutoff \cite{Takayanagi:2011zk}. For the nonadjacent configuration, we take \( \xi_1' = -\infty \), \( \xi_2'=-2R-d\), \( \xi_3'=-2R\), and \( \xi_4' =- \epsilon \to 0 \), as depicted in the right panel of Figure \ref{fig:$BCFT$}. The corresponding cross ratio is $\eta'=\frac{2R}{d}$ and the entanglement entropy becomes
	\begin{equation}
		S_{\text{vN}}(A : B) = \frac{c}{6} \log \left(1+\frac{4R}{d}+2\sqrt{\frac{2R}{d}\left(\frac{2R}{d}+1\right)}\right),
	\end{equation}
	which is again in agreement with equation (\ref{eq: nonadjacent half}).
	
	\bibliographystyle{unsrturl}
	\bibliography{ref202509}

\begin{thebibliography}{10}

\bibitem{Jiang:2025jnk}
Xin Jiang and Haitang Yang.
\newblock {Entanglement Entropy of Conformal Field Theory in All Dimensions}.
\newblock 6 2025.
\newblock \href {http://arxiv.org/abs/2506.02786} {\path{arXiv:2506.02786}}.

\bibitem{Callan:1994py}
Curtis~G. Callan, Jr. and Frank Wilczek.
\newblock {On geometric entropy}.
\newblock {\em Phys. Lett. B}, 333:55--61, 1994.
\newblock \href {http://arxiv.org/abs/hep-th/9401072}
  {\path{arXiv:hep-th/9401072}}, \href
  {http://dx.doi.org/10.1016/0370-2693(94)91007-3}
  {\path{doi:10.1016/0370-2693(94)91007-3}}.

\bibitem{Calabrese:2004eu}
Pasquale Calabrese and John~L. Cardy.
\newblock {Entanglement entropy and quantum field theory}.
\newblock {\em J. Stat. Mech.}, 0406:P06002, 2004.
\newblock \href {http://arxiv.org/abs/hep-th/0405152}
  {\path{arXiv:hep-th/0405152}}, \href
  {http://dx.doi.org/10.1088/1742-5468/2004/06/P06002}
  {\path{doi:10.1088/1742-5468/2004/06/P06002}}.

\bibitem{Ryu:2006bv}
Shinsei Ryu and Tadashi Takayanagi.
\newblock {Holographic derivation of entanglement entropy from AdS/CFT}.
\newblock {\em Phys. Rev. Lett.}, 96:181602, 2006.
\newblock \href {http://arxiv.org/abs/hep-th/0603001}
  {\path{arXiv:hep-th/0603001}}, \href
  {http://dx.doi.org/10.1103/PhysRevLett.96.181602}
  {\path{doi:10.1103/PhysRevLett.96.181602}}.

\bibitem{Ryu:2006ef}
Shinsei Ryu and Tadashi Takayanagi.
\newblock Aspects of holographic entanglement entropy.
\newblock {\em Journal of High Energy Physics}, 2006(08):045--045, aug 2006.
\newblock \href {http://dx.doi.org/10.1088/1126-6708/2006/08/045}
  {\path{doi:10.1088/1126-6708/2006/08/045}}.

\bibitem{Jiang:2024ijx}
Xin Jiang, Peng Wang, Houwen Wu, and Haitang Yang.
\newblock {Alternative to purification in conformal field theory}.
\newblock {\em Phys. Rev. D}, 111(2):L021902, 2025.
\newblock \href {http://arxiv.org/abs/2406.09033} {\path{arXiv:2406.09033}},
  \href {http://dx.doi.org/10.1103/PhysRevD.111.L021902}
  {\path{doi:10.1103/PhysRevD.111.L021902}}.

\bibitem{Jiang:2025tqu}
Xin Jiang, Peng Wang, Houwen Wu, and Haitang Yang.
\newblock {Mixed state entanglement entropy in CFT}.
\newblock {\em JHEP}, 09:133, 2025.
\newblock \href {http://arxiv.org/abs/2501.08198} {\path{arXiv:2501.08198}},
  \href {http://dx.doi.org/10.1007/JHEP09(2025)133}
  {\path{doi:10.1007/JHEP09(2025)133}}.

\bibitem{Jiang:2025dir}
Xin Jiang, Haitang Yang, and Zilin Zhao.
\newblock {Entanglement entropy of mixed state in thermal CFT2}.
\newblock {\em Phys. Rev. D}, 112(4):046025, 2025.
\newblock \href {http://arxiv.org/abs/2501.11302} {\path{arXiv:2501.11302}},
  \href {http://dx.doi.org/10.1103/bpzx-kdgq} {\path{doi:10.1103/bpzx-kdgq}}.

\bibitem{beardon2012geometry}
Alan~F Beardon.
\newblock {\em The geometry of discrete groups}, volume~91.
\newblock Springer Science \& Business Media, 2012.

\bibitem{DiFrancesco:1997nk}
P.~Di~Francesco, P.~Mathieu, and D.~Senechal.
\newblock {\em {Conformal Field Theory}}.
\newblock Graduate Texts in Contemporary Physics. Springer-Verlag, New York,
  1997.
\newblock \href {http://dx.doi.org/10.1007/978-1-4612-2256-9}
  {\path{doi:10.1007/978-1-4612-2256-9}}.

\bibitem{Affleck:1991tk}
Ian Affleck and Andreas W.~W. Ludwig.
\newblock {Universal noninteger 'ground state degeneracy' in critical quantum
  systems}.
\newblock {\em Phys. Rev. Lett.}, 67:161--164, 1991.
\newblock \href {http://dx.doi.org/10.1103/PhysRevLett.67.161}
  {\path{doi:10.1103/PhysRevLett.67.161}}.

\bibitem{Takayanagi:2011zk}
Tadashi Takayanagi.
\newblock {Holographic Dual of BCFT}.
\newblock {\em Phys. Rev. Lett.}, 107:101602, 2011.
\newblock \href {http://arxiv.org/abs/1105.5165} {\path{arXiv:1105.5165}},
  \href {http://dx.doi.org/10.1103/PhysRevLett.107.101602}
  {\path{doi:10.1103/PhysRevLett.107.101602}}.

\end{thebibliography}
\end{document}